\documentclass[a4paper,11pt]{article}

\usepackage[top=2.5cm,bottom=2.5cm, left=2.5cm, right=2.5cm]{geometry}
\usepackage{amsmath,amssymb,amsthm}
\usepackage{amsfonts}
\usepackage{graphicx}
\usepackage{bm}
\usepackage[utf8]{inputenc}
\usepackage[T1]{fontenc}
\usepackage{enumerate}
\usepackage{color}
\usepackage[sort,comma]{natbib}
\usepackage{float}
\usepackage[colorlinks=true, allcolors=blue]{hyperref}
\usepackage{changepage}
\usepackage{dirtytalk}
\usepackage{pdflscape}
\usepackage{booktabs}

\usepackage{ulem}
\usepackage{etoolbox}
\usepackage{tikz}

\newrobustcmd*{\mysquare}[1]{\tikz{\filldraw[draw=#1,fill=#1] (0,0)
rectangle (0.2cm,0.2cm);}}
\newrobustcmd*{\mycircle}[1]{\tikz{\filldraw[draw=#1,fill=#1] (0,0) circle [radius=0.1cm];}}
\newrobustcmd*{\mytriangle}[1]{\tikz{\filldraw[draw=#1,fill=#1] (0,0) --
(0.2cm,0) -- (0.1cm,0.2cm);}}

\newtheorem{theorem}{Theorem}[section]

\theoremstyle{definition}
\newtheorem{definition}[theorem]{Definition}

\theoremstyle{remark}

\title{A Jensen-Shannon divergence based $k$--$NN$ algorithm for missing value imputation in compositional data}
\author{Michail Tsagris, Connie Stewart and Abdulaziz Alenazi \\
\\
Department of Economics, University of Crete, Greece,  \\
\href{mailto:mtsagris@uoc.gr}{mtsagris@uoc.gr} \\
Department of Mathematics and Statistics, University of New Brunswick, Saint John, Canada \\ \href{mailto:connie.stewart@unb.ca}{connie.stewart@unb.ca} \\ 
Department of Mathematics, College of Science \& Center for \\
Scientific Research and Entrepreneurship, Northern Border University, Arar, Saudi Arabia, \\ \href{mailto:a.alenazi@nbu.edu.sa}{a.alenazi@nbu.edu.sa} \\
}

\begin{document}

\maketitle

\begin{center}
{\bf Abstract}
\end{center}
A novel nonparametric method to impute missing values in compositional data is developed. The method is based on the $k$--$NN$ algorithm, utilizes the Jensen-Shannon divergence and employs the Fr{\'e}chet mean to allow for more flexibility in the estimation process. As an extra feature, the hyper-parameters can be self-adaptive according to the pattern of missing values. Unlike restrictive parametric models, the proposed method makes no assumption about the structure of the data and, most importantly, it is applicable even when compositional data contain zero values. Through simulation studies using real data, it is shown that the proposed algorithm outperforms competing algorithms at various settings, not only in terms of accuracy but also in terms of computational efficiency.  \\
\\
\textbf{Keywords}: compositional data, missing values, $k$--$NN$ algorithm, Fr{\'e}chet mean

\section{Introduction}
Missing values in real-life data are common and can be divided into three categories \citep{little2019}. Values in a data set are missing completely at random (MCAR) if the events that lead to any particular data-item being missing are independent both of observable variables and of unobservable parameters of interest, and occur entirely at random. Missing at random (MAR) occurs when the missingness is not random, but instead  can be fully accounted for by variables where there is complete information. Lastly, missing not at random (MNAR) (also known as non-ignorable non-response) involves data that are neither MCAR nor MAR (i.e. the value of the variable that is missing is related to the reason it is missing). This paper addresses the imputation of missing values under the MAR case scenario, with an extension that can be used when there are patterns in the missing values.

 Numerous methods have been developed for imputing missing values in multivariate data, but significantly fewer approaches have been tailored to address the unique challenges posed by compositional data.  Compositional data are non-negative multivariate vectors that convey only relative information, often normalized to sum to 1, and the corresponding treatment of missing values must take into account their restrictive sample space.  Specifically, the sample space is the standard simplex given below
\begin{eqnarray} \label{simplex}
\mathbb{S}^{D-1}=\left\lbrace(u_1,...,u_D)^\top \bigg\vert u_i \geq 0,\sum_{i=1}^Du_i=1\right\rbrace, 
\end{eqnarray}
where $D$ denotes the number of variables, usually referred to as components. 

Compositional data are prevalent across numerous application domains (see \cite{tsagris2020} for a variety of examples), and there is a wide range of literature published on methodology for properly analyzing compositional data. The traditional recommended approach for handling compositional data involves transforming the data to Euclidean space using a log-ratio transformation, followed by the application of standard multivariate techniques.  Consequently, zeros in compositional data pose challenges for this strategy, and there has been substantial interest in alternative techniques for dealing with compositional data containing zeros. 

The literature differentiates between the types of zeros. Structural or essential zeros refer to zeros representing a true absence in the components, while rounded zeros refer to components that have been either rounded to zero or fall below detection limit. For instance, in ecology, the diet composition of predators may be estimated using stomach content analysis. A structural zero can occur if a particular species is not found in the contents. Zeros in fatty acid signatures, which are also used to estimate diet, can be due to limitations of the measuring device and are typically treated as rounded zeros (\cite{stewart2014}). Depending on the type of zeros present, different methodologies are recommended, but dealing with structural zeros is generally considered to be more challenging. Approaches for handling this type of zero have been proposed in a variety of contexts (\cite{stewart2011}, \cite{tsagris2018b}, \cite{scealy2011}, \cite{bear2016}). We note that as our proposed procedure does not use log-ratio transformations, it is applicable when either type of zero, including structural, is present in the data.


\cite{ait2003} proposed using zero value substitution or imputation strategies for replacing rounded zeros occurring in compositional data while \cite{martin2003b} proposed nonparametric imputation to tackle the problem of either rounded zeros or missing values. The literature documents more broadly strategies for the replacement of rounded zeros, compared to methods for imputing general missing values. However, in some cases, the proposed algorithms for imputing rounded zeros can be adjusted to accommodate missing values \citep{palarea2008}. 

In contrast, \cite{hron2010} suggested two algorithms for imputing missing values for compositional data which encompass both classical and robust versions.  The first algorithm imputes missing values using the $k$--$NN$ procedure and the Aitchison's distance (\ref{aitdist}), a  commonly accepted distance measure in compositional data analysis. With this algorithm, a missing value in a compositional vector is imputed using the observed component values of the vector and computing the Aitchison's distance from all other compositional vectors that have no missing values (using the same components). Based on these distances the $k$ nearest neighbouring vectors are detected. Each component value of those vectors is weighted based on their values and the imputation of the missing component values takes place using the median of those weighted values. As this algorithm does not fully utilize the relationship between the compositional components, \cite{hron2010} also proposed using regression models (linear regression or least trimmed regression) for the imputation with the models being applied in an iterative fashion until some convergence criteria based on the covariance matrix of the log-ratio transformed data is met. Both of these algorithms are available in the \textit{R} package \textit{robCompositions} \citep{robcompositions2011}.

Alternative strategies include the multiple imputation by chained equations (MICE) framework that offers a flexible framework for handling missing data in complex datasets \citep{vanBuuren2006, vanBuuren2011}. Non-parametric imputation using random forest \citep{doove2014, shah2014, xiaoqin2017} is another option. A limitation of these methods is the requirement to apply log‑ratio transformations prior to analysis, which precludes the presence of zeros in the data. While zero value imputations could be applied, the modifications can induce bias \citep{tsagris2015a}, especially when the zeros are structural and not rounded.

This paper proposes a new $k$--$NN$ based algorithm for missing value imputation in compositional data that uses the Jensen-Shannon divergence (JSD) and the flexible sample Fr{\'e}chet mean (\ref{frechet}) defined by \cite{tsagris2011}. The Fr{\'e}chet mean extends the simple arithmetic mean by introducing a power hyper-parameter, $\alpha$. The proposed algorithm  offers several advantages over the competing methods in \cite{hron2010}. In particular, our proposed algorithm allows for zeros in the non-missing parts and is also more computationally efficient in terms of tuning the value of $k$ and imputing the missing values. An extension to the algorithm is also developed that allows the hyper-parameters ($\alpha$ and $k$) to be self-adaptive, depending on the pattern of missing values. Simulation studies based on several examples of real-life data demonstrate that the proposed algorithms have wide-ranging applicability and are consistently more accurate than the methods in \cite{hron2010}.   

Disaggregation of multivariate data can yield missing information. For example, \cite{xavier2018} dis-aggregated agricultural data concerning land-use at the detailed pixel level. Concerning the compositional data field specifically, consider the agriculture application where, in some geographical areas, information is provided on the production of numerous crops, whereas in other areas production of groups of crops is reported, and the task of interest is to dis-aggregate the grouped crop production to match other areas. The need to dis-aggregrate data (and estimate missing values) is encountered in other contexts as well. For instance, while many countries report the total number of deaths, some may not provide the number of deaths attributed to specific causes \citep{lopez2020}. 

The new missing value imputation algorithm is introduced in Section \ref{JSKNN}. Extensive simulation studies are displayed in Section \ref{simulations} illustrating the performance of the proposed algorithms using real-life data with varying features. Concluding remarks are outlined in Section \ref{conclusions}.

\section{Missing value imputation in compositional data} \label{JSKNN}
The proposed algorithm and extensions impute missing values arising in compositional data using the $k$--$NN$ procedure which, in turn, requires measuring distance between compositional vectors. Measuring distance in the simplex necessitates special considerations and Euclidean distance is generally regarded as unsuitable in this context due to it not satisfying key properties described below. In the following, we first formally define the types of missing values and then contrast Aitchison's distance measure, previously used for missing value imputation in compositional data \cite{hron2010}, and the JSD which underpins our proposed method detailed in Subsection \ref{knn}. 

\subsection{Formal missing values conditions}

\begin{definition} \label{def:mcar}
Let $\mathbf{R} = (R_1, \ldots, R_D)$ be the missingness indicator vector, where $R_j = 1$ if component $j$ is observed and $R_j = 0$ if missing. The missing data are \textbf{MCAR} if:
\begin{equation}
P(\mathbf{R} = \mathbf{r} \mid \mathbf{X} = \mathbf{x}) = P(\mathbf{R} = \mathbf{r})
\end{equation}
for all $\mathbf{x} \in \mathcal{S}^{D-1}$ and all missingness patterns $\mathbf{r} \in \{0,1\}^D$. \end{definition}

\begin{definition} \label{def:mar}
The missing data are \textbf{MAR} if:
\begin{equation}
P(\mathbf{R} = \mathbf{r} \mid \mathbf{X} = \mathbf{x}) = P(\mathbf{R} = \mathbf{r} \mid \mathbf{X}^o = \mathbf{x}^o)
\end{equation}
where $\mathbf{X}^o$ denotes the observed components. That is, the probability of missingness depends only on the observed values, not on the missing values.
\end{definition}

MNAR is the most complex type of missing data as the probability of missingness in this case is related to the unobserved data. In this work, we assume the MAR scenario. While MAR cannot be directly tested from observed data alone \citep{molenberghs2007}, some diagnostic approaches can be used to assess plausibility. For example, the distributions of the observed components,  $\mathbf{X}^o$, between complete and incomplete observations may be compared with large differences suggesting potential MNAR. However, with compositional data, we could examine whether missingness is related to component magnitudes (for example, small values being more likely missing suggesting censoring MNAR) or check specific component combinations which may indicate informative missingness. External validation could also be performed via a subset with complete data to assess imputation accuracy.




\subsection{Distance Measures for Compositional Data}

\subsubsection{Aitchison's distance}
\cite{ait2003} defined distance between two compositional vectors $\bf x$ and $\bf y$ as
\begin{eqnarray} \label{aitdist}
d\left({\bf x}, {\bf y}\right)=\left[\sum_{j=1}^D\left(\log{\frac{x_j}{g({\bf x})}} - \log{\frac{y_j}{g({\bf y})}}\right)^2 \right]^{1/2},
\end{eqnarray}
where 
$g\left({\bf x} \right)=\prod_{j=1}^Dx_j^{1/D}$ is the geometric mean of the components of $\mathbf{x}$ and $\log{(.)}$ refers to the natural logarithm (logarithm with base $e$). Alternatively, the Aitchison's distance between two compositions is the ordinary Euclidean distance between the compositions transformed by the centered logratio (clr) transformation (\cite{ait2003}). Aitchison's distance is the key component of Hron et al.'s (2010) $k$--$NN$ algorithm. 

\citet{ait1992} argued that a distance on the simplex should satisfy certain properties, namely scale invariance, sub-compositional dominance and perturbation invariance.  Scale invariance ensures that the distance between $\mathbf{x}$ and $\mathbf{y}$ is the same as the distance between $k\mathbf{x}$ and $K\mathbf{y}$, for positive constants $k$ and $K$. Concerning subcompositional dominance, consider two compositional vectors and select sub-vectors, each consisting of the same components. Sub-compositional dominance means that the distance between the sub-vectors is always less than or equal to the distance between the original compositional vectors.  Finally, the perturbation requirement is that the distance between compositional vectors ${\bf x}$ and ${\bf y}$ should be the same as distance between ${\bf x}\oplus_0{\bf p}$ and ${\bf y}\oplus_0{\bf p}$, where the operator $\oplus_0$ means element-wise multiplication and then division by the sum, so that the resulting vectors belong to $S^d$, and ${\bf p}$ is any vector (not necessarily compositional) with positive components.

In contrast, \cite{scealy2014} argued that these properties were derived from the attributes of the log-ratio methods themselves, and then used to justify those same methods as uniquely valid, therefore reversing proper scientific reasoning. Despite design intent, log-ratio methods violate subcompositional coherence when, for example, analyzing sub-partitioned data or when applying robust methods (since outliers may be masked in the full composition but not in some subcomposition). Specifically for the subcompositional coherence property, we quote \cite{scealy2014} "\textit{Moreover, it is applied selectively because it is not actually satisfied by the log-ratio methods it is intended to justify.}" Further, when problems do not allow for log-ratio methods (due to the presence of zeros, for example), these data have been defined as "non-compositional" rather than acknowledging those cases as limitations. Regardless, from a practical point of view, when compositional vectors are normalized prior to analysis, as is assumed here, the scale invariant property is irrelevant. With respect to distance measures, when compositional data contain zeros,  reasonable trade-offs may be necessary \citep{stewart2017}, as Aitchison's distance cannot accommodate zeros directly.


\subsubsection{Jensen-Shannon divergence}
Our proposed algorithm employs an alternate approach for measuring distance between two compositional vectors $\bf x$ and ${\bf y} \in \mathbb{S}^{D-1}$, namely the Jensen-Shannon divergence (JSD) (multiplied by a factor of 2)
\begin{eqnarray} \label{js} 
\text{JSD}\left({\bf x},{\bf y}\right)=\sum_{j=1}^D\left( x_j\log{\frac{2x_j}{x_j+y_j}}+y_j\log{\frac{2y_j}{x_j+y_j}} \right).
\end{eqnarray}

Note that JSD is bounded from above by $2\log 2$. \cite{endres2003} and \cite{osterreicher2003} independently proved that the square root of (\ref{js}), $\sqrt{\text{JSD}}$, satisfies the triangular identity and thus it is a metric\footnote{For a series of inequalities see \cite{lin1991}.}. Moreover, \cite{endres2003} showed that the limiting behavior of JSD as ${\bf x} \rightarrow {\bf y}$ is approximately equal to the $\chi^2$ distance, that is $\text{JSD}\left({\bf x},{\bf y}\right) \approx \sum_{j=1}^D\frac{1}{4y_j}\left(x_j-y_j\right)^2$. 

A useful practical property of the JSD is that, unlike the Aitchison distance or the Kullback-Leibler divergence (KLD), an alternative measure of difference in compositional data, zero values are treated naturally since $0\log0=0$, thus allowing the application of the proposed imputation algorithm to compositional data with zeros. JSD is a member of the $\phi$-divergence family and can be seen as a symmetrized version of the KLD
\begin{eqnarray*}
\text{JSD}\left({\bf x},{\bf y}\right) = KL\left({\bf x},{\bf M}\right) + KL\left({\bf y},{\bf M}\right)=\sum_{j=1}^D\left( x_j\log{\frac{x_j}{M_j}}+y_j\log{\frac{y_j}{M_j}}\right),
\end{eqnarray*}
where ${\bf M}=\frac{{\bf x}+{\bf y}}{2}$. 

\subsubsection{Contour plots of Aitchison's distance and JSD}
\begin{figure}[h!]
\centering
\begin{tabular}{cc}
\includegraphics[scale = 0.4]{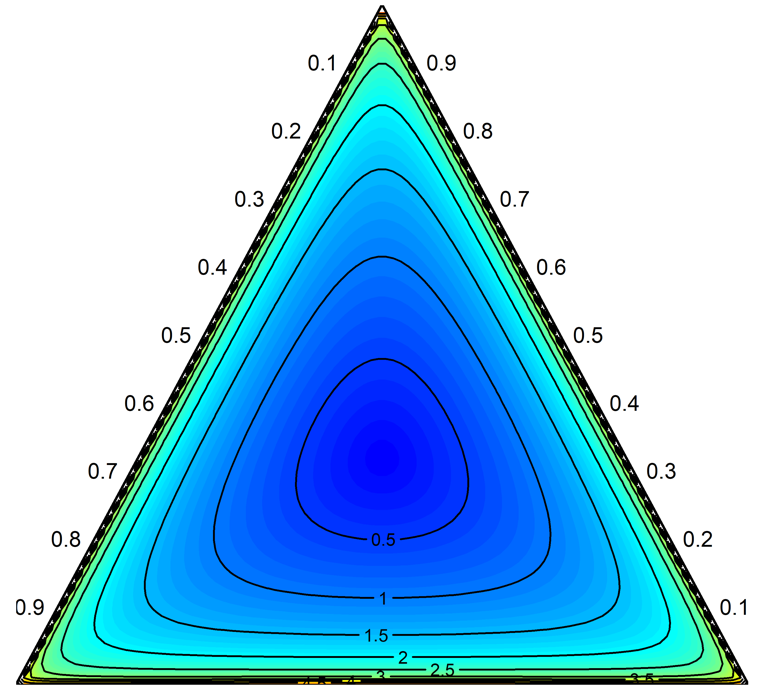}  &
\includegraphics[scale = 0.4]{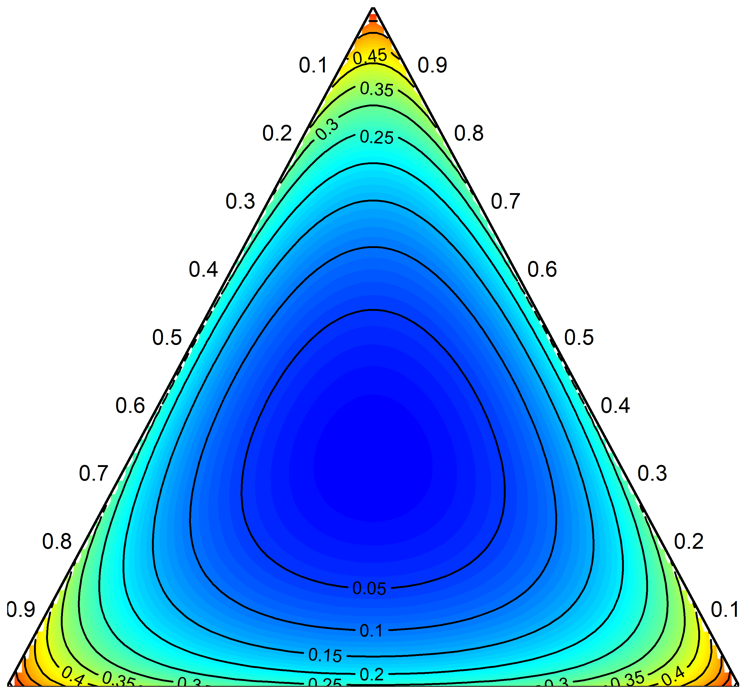}  \\
(a) Aitchison's distance  &  (b) JSD
\end{tabular}
\caption{Contour plots of the Aitchison's distance and the Jensen-Shannon Divergence. . \label{contours} }
\end{figure}

Contour plots visualizing the Aitchison's distance (\ref{aitdist}) and the JSD (\ref{js}) illustrate the differences between these two distances. The produced contour plots show both simplicial distances for many points on the $\mathbb{S}^2$ from its centre. Evidently Aitchison's distance produces contours (Figure \ref{contours}(a)) that better fit the triangle (ternary plot) compared to the contours produced by JSD (Figure \ref{contours}(b)) that are more circular. However, while the Aitchison's distance contours may be perceived as more advantageous compared to the JSD's contours (their shape seems to fit better in the triangle as seen in Figure \ref{contours}(a)), JSD's more circular contours do not appear to affect the proposed algorithm's performance in practice as we show in Section \ref{simulations}.

\subsection{The JSD based $k$--$NN$ algorithm } 
\label{knn}

\subsubsection{Basic Algorithm}
In general terms, the $k$--$NN$ algorithm is a nonparametric technique used for predictions. While it can be computationally heavy, it is straightforward to implement. The procedure relies solely on distance between points and makes no parametric assumptions about the data. Care must be taken when applying the algorithm to compositional data, and in particular when missing values are present. The peculiar structure of the simplex space (\ref{simplex}) could yield inconsistent values if the missing-free variables are used in the same manner as the case with Euclidean multivariate data since it is advisable to utilize the information from all variables of the missing-free compositional vectors. Further, since structural zero values might be present in those vectors, we prefer measuring distance using JSD over Aitchison's distance. 

The proposed algorithm, which we name JSD $k$--$NN$, is conceptually similar to the (1)(a) approach of \citep{hron2010} as it searches the $k$ nearest neighbours among all complete observations. JSD $k$--$NN$ consists of 6 steps, presented below and accompanied by an example compositional data set $\bf X$. In the steps below, $M$ and $O$ subscripts refer to rows of $\bf X$ that contain missing values or no missing values respectively.  The superscript $o$ is used to denote observed components, $m$ refers to missing components and $c$ denotes that components have been normalized.

\begin{enumerate}
\item[Step 1:] Select the value of $k$, the number of nearest neighbours. Automating the choice of $k$ through cross-validation is described in Section \ref{tuning}.
\item[Step 2:] For a compositional data set, ${\bf X}$, separate the compositional data with missing values, ${\bf X}_M$, from the complete/observed compositional data ${\bf X}_O$. 
\ \\
\ \\
As an example, let 
\[{\bf X} = \begin{bmatrix}
            \bm{0.2} & \textit{NA}  & \bm{0.3} & \bm{0.1} & \textit{NA} \\
            \bm{0.1} & \textit{0.2} & \bm{0.4} & \bm{0.1} & \textit{0.2} \\
            \bm{0.2} & \textit{0.4} & \bm{0.2} & \bm{0.1} & \textit{0.1} \\
            \bm{0.1} & \textit{0.3} & \bm{0.3} & \bm{0.2} & \textit{0.1} \\
\end{bmatrix},\]
then 
\[{\bf X}_M = \begin{bmatrix}
            \bm{0.2} & \textit{NA}  & \bm{0.3} & \bm{0.1} & \textit{NA} \\
\end{bmatrix}
\]
and
\[{\bf X}_O = \begin{bmatrix}
            \bm{0.1} & \textit{0.2} & \bm{0.4} & \bm{0.1} & \textit{0.2} \\
            \bm{0.2} & \textit{0.4} & \bm{0.2} & \bm{0.1} & \textit{0.1} \\
            \bm{0.1} & \textit{0.3} & \bm{0.3} & \bm{0.2} & \textit{0.1} \\
\end{bmatrix}.\]

\item[Step 3a:] Denote the $i$th compositional vector of observed (non-missing) values in 
${\bf X}_M$ by ${\bf X}^o_{M,i}$ and the same components in the rows of ${\bf X}_O$ by ${\bf X}^o_O$.  The remaining components of ${\bf X}_O$ are denoted by ${\bf X}^m_O$. From Step 2, we will separate the observed data that lie in columns 1, 3 and 4:
\[{\bf X}_{M,1}^o = \begin{bmatrix}
            \bm{0.2} & \bm{0.3} & \bm{0.1} \\
\end{bmatrix},\]
\[{\bf X}_O^o = \begin{bmatrix}
            \bm{0.1} & \bm{0.4} & \bm{0.1} \\
            \bm{0.2} & \bm{0.2} & \bm{0.1}  \\
            \bm{0.1} & \bm{0.3} & \bm{0.2}  \\
\end{bmatrix}\]
and the observed data that lie in columns 2 and 5:
\[{\bf X}_O^m = \begin{bmatrix}
            \textit{0.2} & \textit{0.2} \\
            \textit{0.4} & \textit{0.1} \\
            \textit{0.3} & \textit{0.1} \\
\end{bmatrix}.\]
\item[Step 3b:]
Normalize the rows of ${\bf X}^o_{M,i}$ and ${\bf X}^o_O$ and denote these by  ${\bf X}^{(c, o)}_{M,i}$ and ${\bf X}^{(c, o)}_O$ respectively. That is, ${\bf X}^{(c, o)}_{M,i}$ refers to the normalized observed components of the $i$th incomplete vector, and ${\bf X}^{(c, o)}_O$ similarly denotes the same normalized components, but of those rows containing complete vectors. Also, denote the difference from unity of the sum of the values in ${\bf X}^o_{M,i}$ by $T_i$. That is,  $T_i=1-\sum_{j=1}^D{\bf X}^{o,j}_{M,i}$ and represents the total of the non-missing components for the $ith$ incomplete vector.

From Step 3a, 
\[{\bf X}_{M,1}^{(c,o)} = {\bf X}_{M,1}^{(o)}/T_1 =  
\begin{bmatrix}
            \bm{0.2} & \bm{0.3} & \bm{0.1} \\
\end{bmatrix} / (0.2+0.3+0.1) = 
\begin{bmatrix}
            0.33 & 0.50 & 0.17 \\
\end{bmatrix},
\]
\[{\bf X}_O^{(c,o)} = {\bf X}_O^{(o)} / (\bm{1} - \bm{T}) = 
\begin{bmatrix}
            (\bm{0.1} & \bm{0.4} & \bm{0.1})/(0.1+0.4+0.1) \\
            (\bm{0.2} & \bm{0.2} & \bm{0.1})/(0.2+0.2+0.1)  \\
            (\bm{0.1} & \bm{0.3} & \bm{0.2})/(0.1+0.3+0.2)  \\
\end{bmatrix} = 
\begin{bmatrix}
            0.17 & 0.67 & 0.17 \\
            0.40 & 0.40 & 0.20  \\
            0.17 & 0.50 & 0.33  \\
\end{bmatrix}.\]
\item[Step 4:] Compute the JSD is  between ${\bf X}^{(c,o)}_{M,i}$ and every vector in ${\bf X}^{(c,o)}_O$. 

\ \\
\ \\
In Step 3b, ${\bf X}_{M,1}^{(c,o)} = [0.33, 0.50, 0.17]$ so we compute
\begin{eqnarray*}
    \mathrm{JSD}\left ( [0.33 , 0.50 , 0.17],[0.17 , 0.67 , 0.17] \right ) & = & 0.039 \nonumber \\
    \mathrm{JSD}\left ( [0.33 , 0.50 , 0.17],[0.40 , 0.40 , 0.20] \right )  & = & 0.010 \nonumber\\
    \mathrm{JSD}\left ( [0.33 , 0.50 , 0.17],[0.17 , 0.50 , 0.33] \right ) & = & 0.052 \nonumber     
\end{eqnarray*} 
\item[Step 5:] Select the $k$ compositional vectors in ${\bf X}_O$ that correspond to the $k$ smallest JSDs.  Denote the arithmetic mean of these vectors by $\hat{\bm{\mu}}_{O,k,i}$. 

If, for example, $k=2$ then the 2 smallest distances in Step 4 (that is, 0.039 and 0.010) were obtained using $[0.17 ,0.67, 0.17]$ and $[0.40 , 0.40 , 0.20]$ respectively and these are subcompositions of rows 1 and 2 of ${\bf X}_O$.

Averaging rows 1 and 2 of ${\bf X}_O$ yields
\begin{eqnarray*}
    \hat{\bm{\mu}}_{O,2,1} & = & \frac{1}{2}\left [ (0.1,\textit{0.2},0.4,0.1,\textit{0.2}) + (0.2,\textit{0.4},0.2,0.1,\textit{0.1}) \right ] \\
                     & = & (0.15, \textit{0.30}, 0.30, 0.10, \textit{0.15}).
\end{eqnarray*}
\item[Step 6:] Normalize the components of $\hat{\bm{\mu}}_{O,k,i}$ that correspond to the missing components in the $i$th row of ${\bf X}_{M}$. Denote this value by $\hat{\bm{\mu}}^m_{O,k,i}$. Multiply the missing values by $T_i$ and use these to impute the missing values in ${\bf X}_{M,i}$. That is ${\bf \hat{X}}^m_{M,i}=\hat{\bm{\mu}}^{(c,m)}_{O,k,i} T_i$ and is obtained from the  arithmetic average of the $k$ nearest neighbors in Step 5 (normalized and then weighted) of only those proportions in the complete vectors corresponding to the positions of missing components in the $i$th incomplete vector.

Components in positions 2 and 5 of the first row of ${\bf X}_{M}$ in Step 2 are missing values and these correspond to components 0.3 and 0.15 respectively in $\hat{\bm{\mu}}_{O,2,1}$ in Step 5. Therefore, 
\[\hat{\bm{\mu}}_{O,2,1}^m = (0.3,0.15) \Rightarrow \hat{\bm{\mu}}^{(c,m)}_{O,2,1} = (0.3,0.15)/(0.3+0.15)= (0.67,0.33). \]

Since $T_1= 0.4$ (from Step 3b), we multiply the normalized components $\hat{\bm{\mu}}^{(c,m)}_{O,2,1}$ by $T_1$ to obtain
\[ {\bf\hat{X}}^m_{M,1} = (0.67,0.33)(0.4) = (0.27, 0.13) \]
and the imputed first row of $\bf{X}$ is then 
\[
\begin{bmatrix}
            \bm{0.2} & \textit{NA} & \bm{0.3} & \bm{0.1} & \textit{NA} 
\end{bmatrix} \rightarrow
\begin{bmatrix}
            \bm{0.2} & \textit{0.27} & \bm{0.3} & \bm{0.1} & \textit{0.13} \\
\end{bmatrix}.\]
\end{enumerate}

It is important to note that while the missing values are imputed, the values of the other components remain unaltered. In Section \ref{simulations, extensive simulation studies based on real data illustrate empirically the algorithm's performance and advantages over the competing methods.}



\subsection{JSD $\alpha$--$k$--$NN$ algorithm}
\cite{tsagris2011} defined the sample Fr{\'e}chet mean for compositional data to be
\begin{eqnarray} \label{frechet}
\hat{\bm{\mu}}_{\alpha}\left({\bf x}\right)=\mathcal{C}\left\lbrace \left\lbrace \left(\sum_{i=1}^n \frac{x_{ij}^\alpha}{\sum_{k=1}^Dx_{kj}^\alpha}\right)^{1/\alpha}\right \rbrace_{j=1,\ldots,D} \right \rbrace,
\end{eqnarray}
where $\alpha$ is a tunable hyper-parameter that ranges between $-1$ and $1$ \citep{tsagris2011,tsagris2020} (or between 0 and 1 when zeros are present) and $\mathcal{C}$ denotes the closure (normalization) operation onto the simplex. Two special cases are of particular interest. The Fr{\'e}chet mean (\ref{frechet}) converges to the closed geometric mean, $\hat{\bm{\mu}}_0$ (defined below and in \cite{ait1989}), as $\alpha$ tends to zero 
\begin{eqnarray*}
\lim_{\alpha \rightarrow 0}{\hat{\bm{\mu}}_{\alpha}\left({\bf x}\right)} \rightarrow \hat{\bm{\mu}}_0\left({\bf x}\right) = \mathcal{C}\left\lbrace  \left\lbrace \left(\prod_{j=1}^n x_{ij}\right)^{1/n}  \right \rbrace_{i=1,\ldots,D} \right \rbrace.
\end{eqnarray*}
Additionally, when $\alpha=1$ the Fr{\'e}chet mean is equal to the raw sample arithmetic mean \begin{eqnarray} \label{mean} 
\bm{\mu}_1\left({\bf x}\right)=\left\lbrace\frac{1}{n}\sum_{i=1}^nx_{ij}\right\rbrace_{j=1,\ldots , D}. 
\end{eqnarray}

The Fr{\'e}chet mean has a nice theoretical property. Specifically, \cite{kendall2011} showed that the central limit theorem applies to Fr{\'e}chet means defined on manifold valued data and the simplex space is an example of a manifold \citep{pantazis2019}. Second, for strictly positive compositional data without zero values present, and $\alpha \geq0$, the Fr{\'e}chet meant mean is unique. This is because the function $f(x) = x^\alpha$ is strictly convex on $(0, \infty)$ with appropriate averaging. The power mean with $\alpha \geq 0$ minimizes a strictly convex functional, guaranteeing uniqueness. In the case of $\alpha <0$, when there are no zero values present, the function may not be strictly convex. However, the constraint of the simplex (a compact, convex set) combined with continuity ensures existence. Uniqueness is not guaranteed in general for $\alpha < 0$, though it holds for most practical cases with non-degenerate data.

In the context of missing value imputation the Fr{\'e}chet mean provides a generalization of the raw sample mean and also escapes the log-ratio methodology suggested by \cite{ait2003}. Specifically, the simple arithmetic mean (\ref{mean}) used in Step 5 of the proposed algorithm ($\hat{\bm{\mu}}_{O,2,1}$ in our example) can be substituted by the Fr{\'e}chet mean (\ref{frechet}) to increase the flexibility of the imputations. This modified procedure is referred to as the JSD $\alpha$--$k$--$NN$ algorithm. 

\begin{figure}[h!]
\centering
\includegraphics[scale = 0.5, trim = 0 50 0 0]{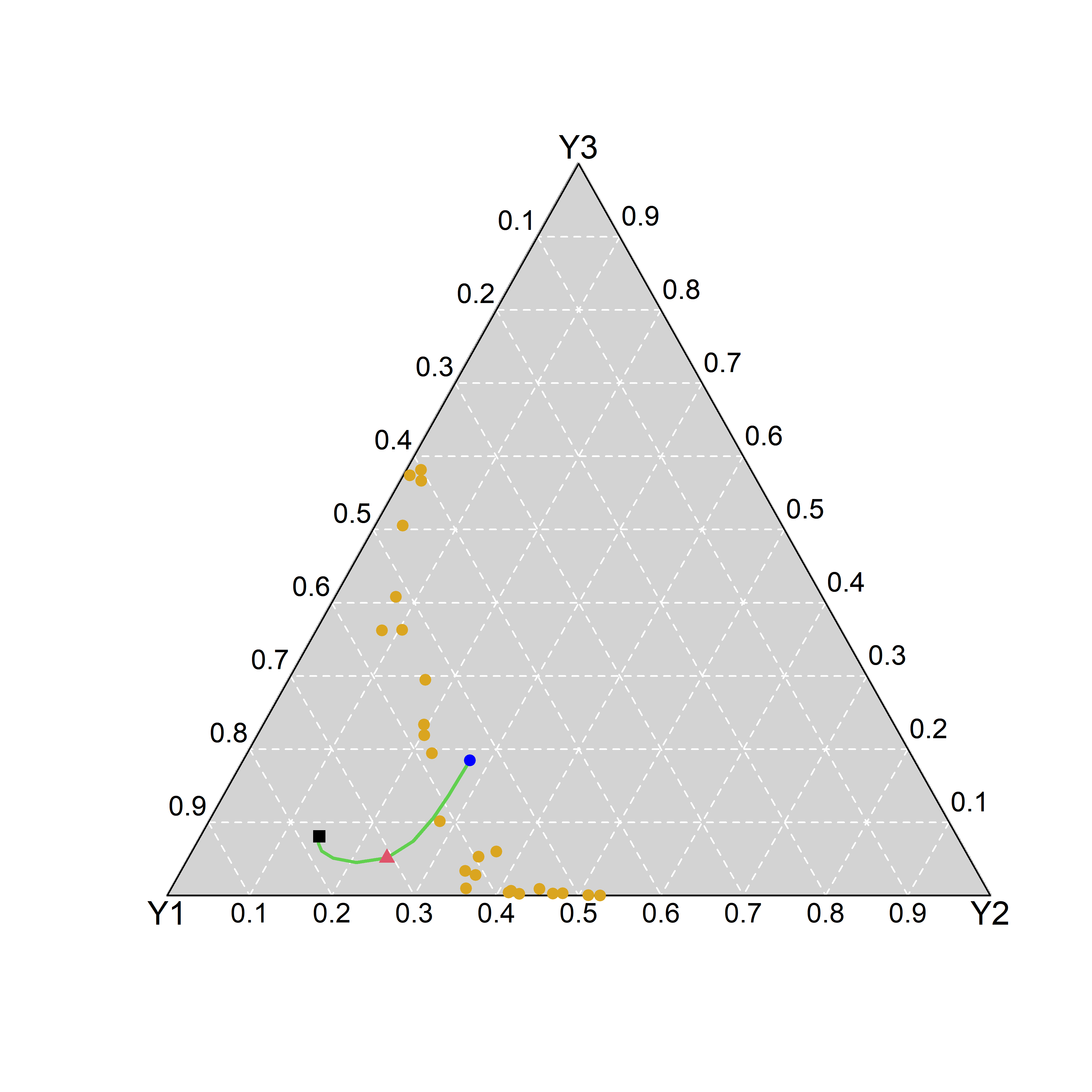} 
\caption{Trajectory of the Fr{\'e}chet mean for a range of values $\alpha$. The symbols are as follows: \mysquare{black} = Fr{\'e}chet mean with $\alpha=-1$, \mytriangle{red} = Fr{\'e}chet mean with $\alpha=0$ and \mycircle{blue} = Fr{\'e}chet mean with $\alpha=1$. The dashed green curve \textbf{-} shows the path of all Fr{\'e}chet means starting with $\alpha=-1$ up to $\alpha=1$. \label{fig_frechet} }
\end{figure}

As an example of the effect of $\alpha$ on the Fr{\'e}chet mean and subsequently on the quality of the imputed values, Figure \ref{fig_frechet} demonstrates the trajectory of the Fr{\'e}chet means for a range of values of $\alpha$ between $-1$ and $1$. Evidently, $\alpha$ has the potential to enhance the quality of the imputations. 

\subsubsection{Tuning the $\alpha$ and $k$ hyper-parameters} \label{tuning}
In the repeated leave-N-out CV protocol, the complete compositional data ${\bf X}_O$ are utilized for tuning the value of $k$, and $N$ is equal to the number of compositional vectors with missing values. To begin, $N$ vectors are randomly selected and filled with missing values (NA) following the observed pattern in ${\bf X}_M$. The proposed algorithm imputes the missing values for a range of values of $\alpha$ and $k$. For each value of $\alpha$ and $k$, Aitchison's distance between the true and imputed compositional vectors in the test data is computed, acting as the performance metric. If zero values are present, the JSD (\ref{js}) is used instead. This procedure is repeated multiple times, and the performance metric of each value of $\alpha$ and $k$ at each repetition is computed from the aggregation of their performances. The pair of values ($\alpha$, $k$) corresponding to the smallest average distance is selected. 

\subsection{Adaptive JSD $\alpha$--$k$--$NN$ algorithm} \label{ada}
In practice, missing values can occur in various and different combinations of components. This implies that one should not use a universal pair of $\alpha$ and $k$ values, but adapt a pair of hyper-parameters values to each pattern. This strategy is evidently computationally heavier and the CV protocol must be applied to each observed pattern separately. The trade-off between added complexity and accuracy is evaluated through simulations.

\section{Simulation studies} \label{simulations}
Four real-life datasets were used in simulation studies\footnote{The simulation studies were performed in the open statistical software \textit{R} 4.4.2. Functions to perform the proposed algorithms exist in the \textit{R} package \textsf{CompositionalNAimp} \citep{tsagris2026}. } to evaluate the algorithms that were introduced in Subsections \ref{knn}-\ref{ada}. These datasets will be referred to as the \textit{Wines}, \citep{hron2010}, \textit{Hydrochemical}  \citep{otero2005}, \textit{Prey Fatty Acids} \citep{Stewart2022} and farm accountancy data network (\textit{FADN}) \citep{mattas2026} datasets. The \textit{Wines} dataset contains information on the concentration (mg/L) of 8 phenolic acids (vanilic, fentisic, protocatechuic, syringic, gallic, coumaric, ferulic and caffeic acid) in 30 Czech wine samples. The \textit{Hydrochemical} dataset consists of 485 observations on 19 hydrochemical variables of the Llobregat river basin water in north-eastern Spain. The \textit{Prey Fatty Acids} contain information on 18 fatty acids of 21 fish species, with 1398 fatty acid signatures (or compositions) in total. Finally, the \textit{FADN} dataset is a subset of the original FADN dataset and refers to a sample of 487 farms conveying information on the production of 10 crops from 4 regions in central Greece. 

The datasets examined possess different attributes that allow us to explore various real-life scenarios.  The \textit{Wines} and \textit{Hydrochemical} datasets do not contain any zeros. The \textit{Prey Fatty Acids} data not only contain zeros, but also are comprised of different classes (species). Lastly, the \textit{FADN} data set, contains zero values, and is an example of disaggregation of components. It is worth noting that the type of zero (rounded or structural) is not relevant as zeros are not modified in the analysis.

Both the JSD $k$--$NN$ and the JSD $\alpha$--$k$--$NN$ algorithms used a range of 2 up to 10 nearest neighbors. When there were no zeros in the dataset, the $\alpha$ values in the JSD $\alpha$--$k$--$NN$  algorithm spanned from -1 up to 1 in increments of 0.1, and the same range of positive values only when the data contained zero values. Cross-validation was used to select the optimal pair of $\alpha$ and $k$, where the metric of performance was Aithchison's distance, in the absence of zero values, or the JSD, when zero values were present.

The specific methods in Hron's et al. (2010) with which are our proposed algorithm is compared are as follows: \textit{rob $k$--$NN$}: the $k$--NN algorithm using Aitchison's distance and the mean (or the median) to aggregate the nearest neighbors, \textit{lm}: least squares regression (on the log-ratio transformed compositional data) within the author's iterative model-based imputation procedure, \textit{ltsReg}: least trimmed squares regression, and \textit{ltsReg2}: least trimmed squares regression (on the log-ratio transformed compositional data) with imputed values perturbed in the direction of the predictor by values drawn form a normal distribution with mean and standard deviation related to the corresponding residuals and multiplied by some noise.

\subsection{\textit{Wines} and \textit{Hydrochemical} Data}
\label{winehydro}

\subsubsection{Case 1: JSD $k$--$NN$ and $\alpha$--$k$--$NN$}
\begin{figure}[h!]
\begin{tabular}{cc}
\includegraphics[scale = 0.4, trim = 50 0 0 0]{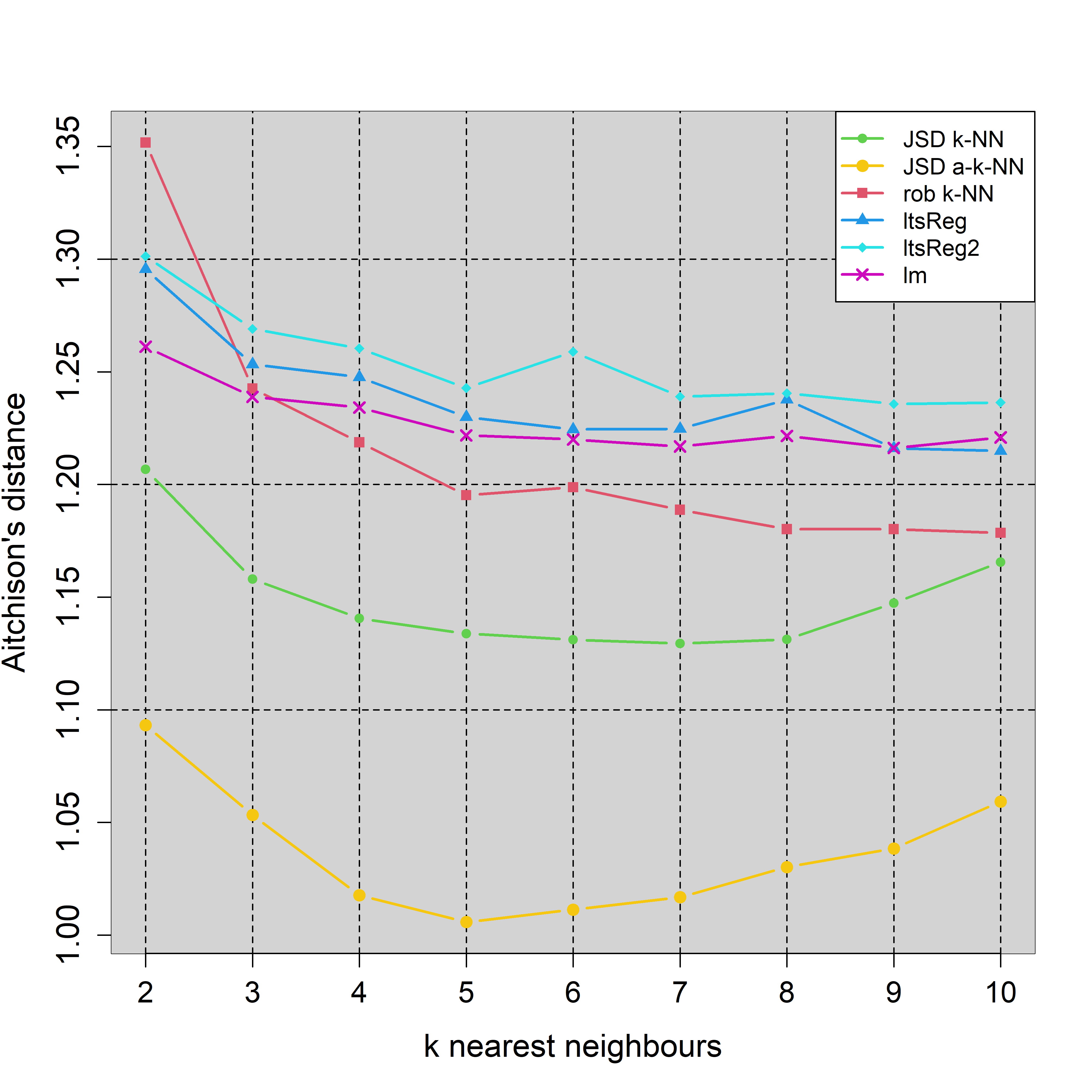}  &
\includegraphics[scale = 0.4, trim = 30 0 0 0]{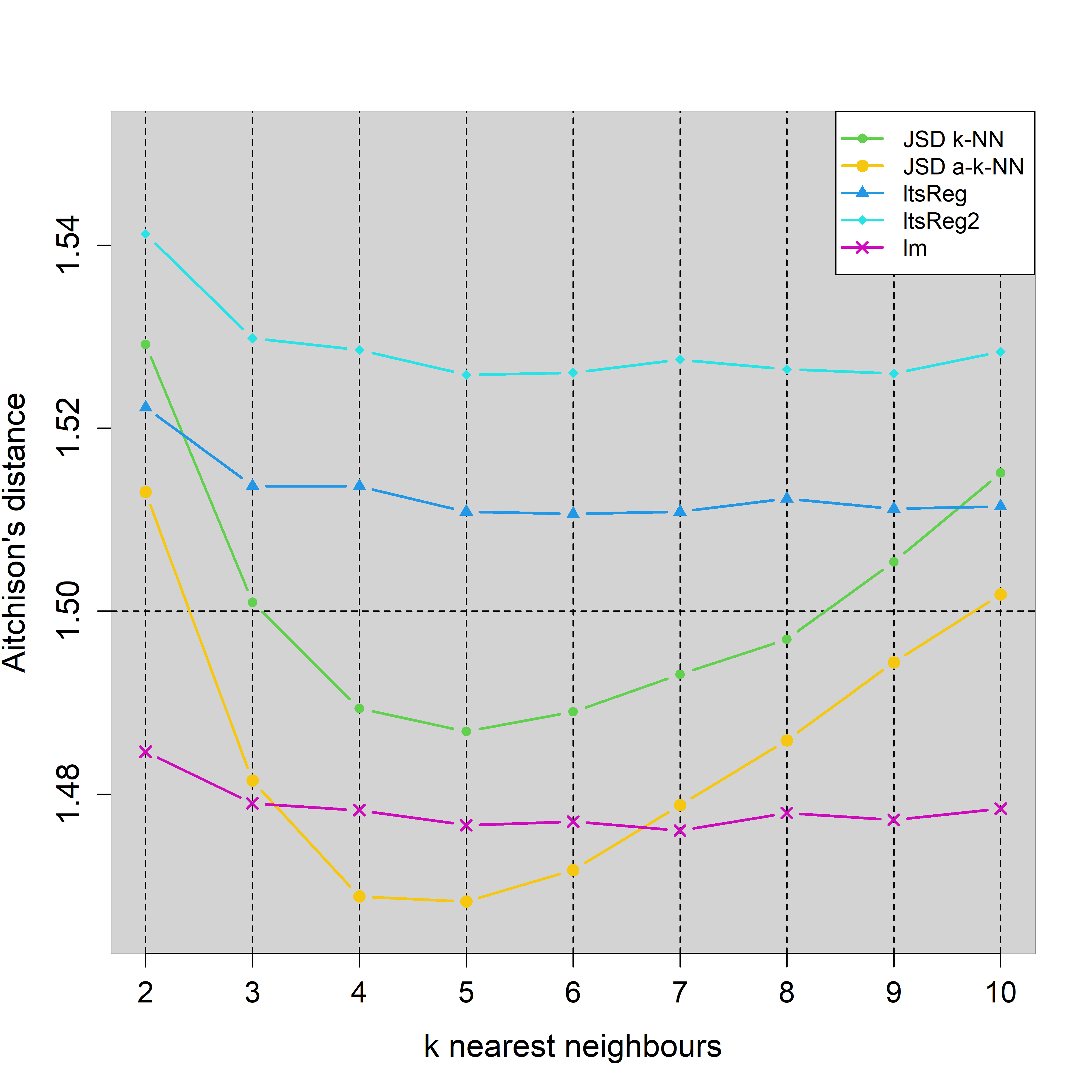}  \\
(c) \textit{Wines} dataset  &  (d) \textit{Hydrochemical} dataset
\end{tabular}
\caption{Average Aitchison's distance between the true and the imputed compositional vectors for a range of nearest neighbours for each method. For clarity of presentation, the results of the rob $k$--$NN$ approach do not appear in the \textit{Hydrochemical} dataset because the resulting Aitchison's distance is too large. \label{fig_real} }
\end{figure}

For both the \textit{Wines} and \textit{Hydrochemical} datasets, 10\% of the compositional vectors were randomly selected and for each of these vectors half of their components were selected and their values substituted with \textit{NA}. Missing values were imputed using the JSD $k$--$NN$ and JSD $\alpha$--$k$--$NN$ procedures, as well as competing methods based on \cite{hron2010} found in the R package \cite{robcompositions2011}. The performance metric was the Aitchison's distance (\ref{aitdist}) between the true and the imputed compositional vectors. The JSD was not employed as a measure of performance as this could entail optimistic bias favoring the proposed algorithm. The process of introducing missing values followed by their imputation was repeated $500$ times. Figure \ref{fig_real} displays the average Aitchison's distances of the JSD $k$--$NN$, the JSD $\alpha$--$k$--$NN$ and of the four other competing algorithms. 

For the JSD $\alpha$--$k$--$NN$ the value of $\alpha$ that yielded the optimal results, on average, was selected for comparison. For the \textit{Wines} dataset, the optimal value of $\alpha$ was equal to 0.5, and the average Aitchison's distance decreased by 0.4\% using the JSD $\alpha$--$k$--$NN$. For the \textit{Hydrochemical} dataset, the optimal value of $\alpha$ was equal to 0.1 and the Aitchison's distance of the JSD $\alpha$--$k$--$NN$ was improved by 0.35\% on average. The JSD $\alpha$--$k$--$NN$ with the optimal values of $\alpha$ and $k$ led to an improvement of 12\% and 2\% over the plain JSD $k$--$NN$, for the Wines and Hydrochemical datasets, respectively. 

\subsubsection{Case 2: Adaptive JSD $\alpha$--$k$--$NN$}
To explore the effectiveness of the adaptive JSD $\alpha$--$k$--$NN$ we changed the scenario of the missing values and each dataset was sorted based on the values of their first component. For the first half of the observations, 10\% of the compositions were randomly selected and then missing values were randomly assigned to half of the components. Another 10\% of the remaining rows were randomly selected and missing values randomly assigned to the other half of the components. Thus, two patterns of missing values were created that changed according to the values of the first component. The Aitchison's distance determined by the optimal values of $\alpha$ and $k$ was compared for the two methods. The adaptive version led to improved performances that ranged between 2\% and 6\%.  

\subsection{Prey Fatty Acids Data}
Recall that the \textit{Prey Fatty Acids} dataset contains zeros and the data are grouped by species. With the JSD $k$--$NN$, the fatty acids that are zero are unchanged during the missing value estimation process and can later be analyzed as either rounded or structural zeros depending on the application \citep{stewart2014}. The simulation scenario laid out in Subsection \ref{winehydro} was used with the \textit{Prey Fatty Acids} to compare the JSD $k$--$NN$, JSD $\alpha$--$k$--$NN$ and adaptive $\alpha$--$k$--$NN$ to the robust $k$--$NN$ algorithm in \citep{hron2010}. However, due to the presence of zero values, the Aitchison's distance (\ref{aitdist}) could not be computed and hence the performance of the algorithms was evaluated using the JSD (\ref{js}). Moreover, only the robust $k$--$NN$ algorithm \citep{hron2010} with the Euclidean distance was used for comparison as the iterative regression based algorithms were not applicable due to the zeros in this dataset.

Figure \ref{fig_real3}(a) shows the average JSD for a range of nearest neighbours. The improvement in the performance of the JSD $k$--$NN$ compared to that of the robust $k$--$NN$ of \cite{hron2010} is 22\%, while the JSD $\alpha$--$k$--$NN$ further improved the JSD $k$--$NN$ by 3.5\%. The adaptive JSD $\alpha$--$k$--$NN$ did not improve the performance of the imputations. 

\subsection{\textit{FADN} Data}
The FADN data provide crop productivity in the Greek NUTS II region of Thessaly during the 2017-2018 cropping year. The data refer to the production of a sample of 487 farms in 10 crops, namely a) other cereals, b) durum wheat, c) maize, d) potatoes, protein crops and rice, e) cotton, f) tobacco, oil seeds, industrial crops and vegetables, g) green plants, pasture and grazing, h) fruits, fruits and nuts, i) olive trees and j) grapes and wine. The four regions of Central Greece, from which the sample data come from, are Karditsa, Larisa, Magnisia and Trikala. To assess the proposed methods, it is supposed that one region reports the production in 8 crops, in the sense that the three cereal crops (Other Cereals, Durum Wheat and Maize) were aggregated into one group of crops. The Trikala region was selected to play this role due to the fact that it had the smallest sample size (73 farms), resulting in 53 farms with missing values\footnote{Some farms did not produce cereals at all.}. 

\begin{figure}[h!]
\begin{tabular}{cc}
\includegraphics[scale = 0.4, trim = 50 0 0 0]{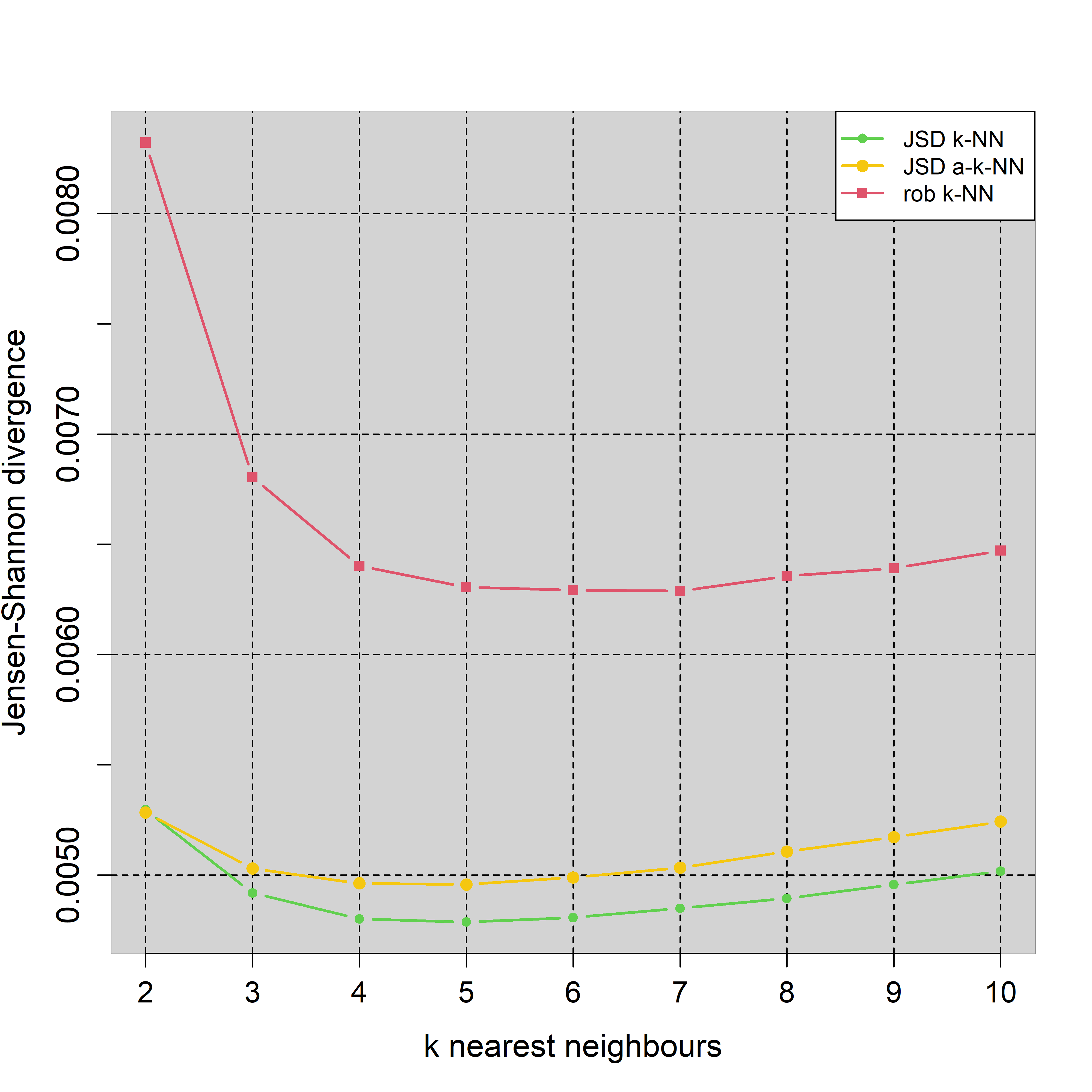}  &
\includegraphics[scale = 0.4, trim = 50 0 0 0]{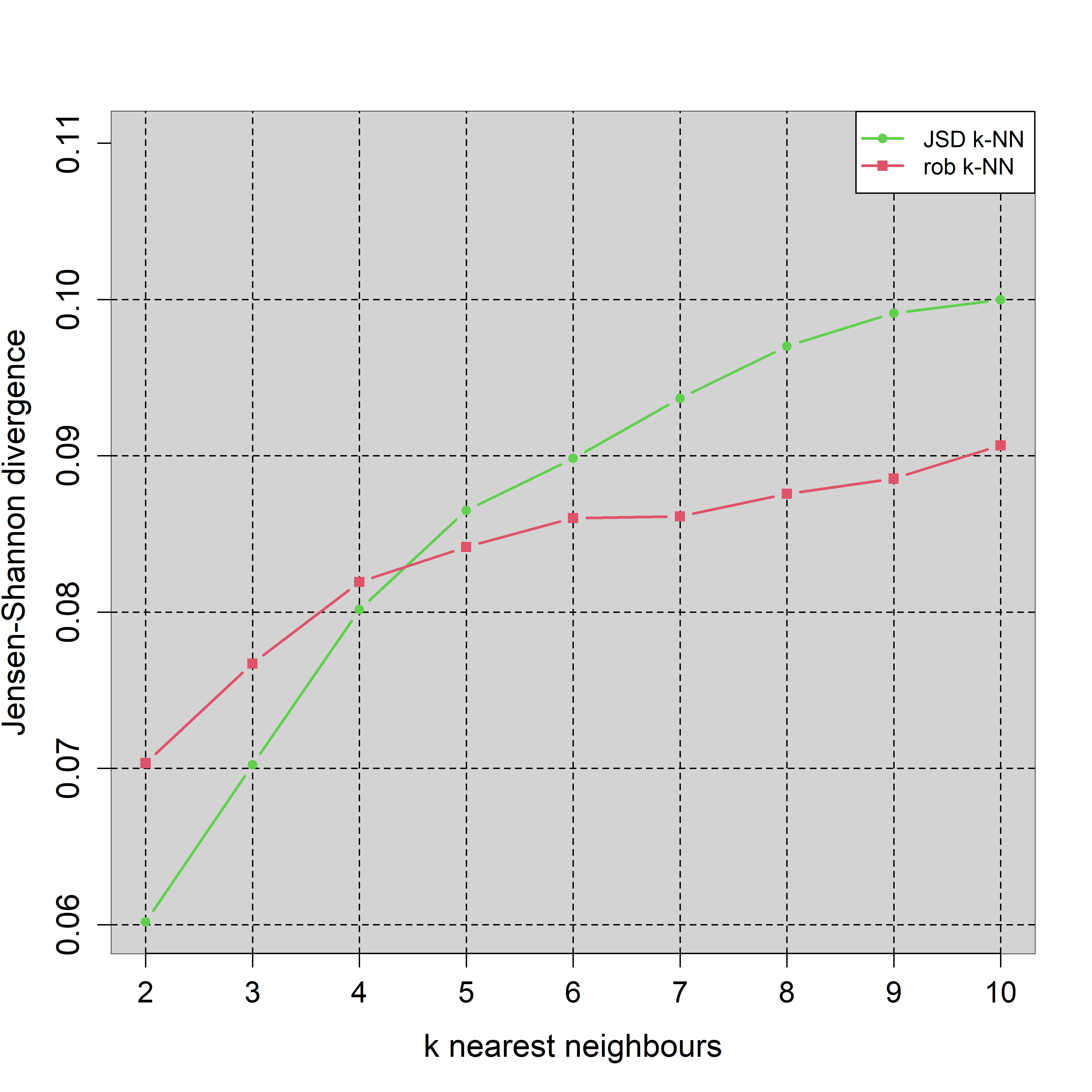} \\
(a) \textit{Prey Fatty Acids} dataset &  (b) \textit{FADN} dataset 
\end{tabular}
\caption{Average JSD between the true and the imputed compositional vectors for a range of nearest neighbours for each method using the \textit{Prey Fatty Acids} (a) and the FADN dataset (b). \label{fig_real3} }
\end{figure}

The objective is to dis-aggregate the total value of the group of cereals into their three components. Due to the fact that this dataset contained farms with zero values, the robust $k$--$NN$ \citep{hron2010} was again the only competitor. Figure \ref{fig_real3}(b) plots the average JSD for a range of nearest neighbours. The JSD $k$--$NN$ outperformed the robust $k$--$NN$, by 15\%, while the JSD $\alpha$--$k$--$NN$ did not further improve the JSD $k$--$NN$ and neither did the adaptive JSD $\alpha$--$k$--$NN$, hence no results are shown for these extensions.

\subsection{Computational cost}
The computational cost to impute the missing values of the JSD $k$--$NN$ and of the robust $k$--$NN$ \citep{hron2010} were measured when a range of nearest neighbours, $k=2,\ldots,10$ were used. Five sample sizes, $n=(500, 1000, 2000, 5000, 10000)$, and 3 dimensionalities, $D=(10, 15, 20)$, were explored. For each combination of sample size and dimensionality, data were generated from the Dirichlet distribution with random parameters\footnote{Defining the parameters is not important as the goal of this experiment is to measure the computational cost of each method}. Ten percent of the simulated vectors were randomly selected and 30\% of their components were set as missing. The two $k$--$NN$ algorithms were run on the simulated data and the computational cost was measured and averaged over 20 repetitions of this process. 

The speed-up factor (ratio of the duration of robust $k$--$NN$ to the duration of JSD $k$--$NN$) appears in Table \ref{times}. Evidently the robust $k$--$NN$ algorithm \citep{hron2010} is computationally heavier compared to the JSD $k$--$NN$. The speed-up factor reduces with increasing sample sizes and increases with increasing dimensionality. 

\begin{table}[ht]
\centering
\caption{Speed-up factors (ratio of the duration of robust $k$--$NN$ to the duration of JSD $k$--$NN$).}
\label{times}
\begin{tabular}{l|rrr}
\toprule
& D=10 & D=15 & D=20 \\ 
\midrule
n=500 & 12.94 & 18.38 & 25.02 \\ 
n=1000 & 11.94 & 16.19 & 23.45 \\ 
n=2000 & 9.88 & 13.14 & 20.02 \\ 
n=5000 & 7.75 & 10.77 & 16.88 \\ 
n=10000 & 5.99 & 8.50 & 13.80 \\ 
\bottomrule
\end{tabular}
\end{table}

\section{Conclusions} \label{conclusions}
In this work, a new technique based on the $k$--$NN$ algorithm is proposed for imputing missing values in compositional data, as well as two extensions to further enhance prediction accuracy. The first generalization uses the Fr{\'e}chet mean with a tunable parameter for added flexibility. Alternative metrics, most of which were listed in \cite{tsagris2014} may be used as well. A further modification was proposed to handle data with patterns in the missing values using an adaptive algorithm that allows for different hyper-parameters according to the missingness patterns. Through a variety of simulation studies using diverse real-life datasets, it was shown that the proposed JSD $k$--$NN$ algorithm can yield substantial improvements over competing methods and, furthermore, unlike available algorithms, the presence of zeros posed no issues. In some cases, additional gains were observed for the JSD $\alpha$--$k$--$NN$ and adaptive JSD $\alpha$--$k$--$NN$ algorithms. Another important advantage of the new procedure is the computational cost that is significantly smaller in comparison to the competing algorithms. 

We acknowledge that the adaptive JSD $\alpha$--$k$--$NN$ does not always substantially improve the performance of the imputations. Indeed, determining whether the added complexity is beneficial was one of the motivations for the simulation study. Possible reasons for this phenomenon may be that there are too many sparse patterns and/or number of missing data. While we are not able to offer a heuristic rule concerning when the adaptive method would perform better, results suggest that perhaps in practice, the extension is not usually worthwhile. It is always possible, however, to analyze the data both ways and assess the performance of each via cross-validation.

We note that compositional data occurring in official statistics \citep{hron2010}, such as tax components or income/expenditure components, frequently exhibit some observations that are MNAR and accurate missing value imputation methods are essential, but we have not assessed our algorithm under these circumstances. \\
\\
\textbf{Acknowledgments}: This work was supported by funding from the Natural Sciences and Engineering Research Council of Canada. The authors would also like to acknowledge the reviewers for their thoughtful and constructive feedback.

\bibliographystyle{apalike}
\bibliography{vivlio}

@article{Stewart2022,
author = {Stewart, Connie and Lang, Shelley L. C. and Iverson, Sara and Bowen, W. Don},
title = {Measuring repeatability of compositional diet estimates: An example using quantitative fatty acid signature analysis},
journal = {Ecology and Evolution},
volume = {12},
number = {10},
pages = {e9428},
doi = {https://doi.org/10.1002/ece3.9428},
year = {2022}
}

@book{ait2003,
  title={The statistical analysis of compositional data},
  author={Aitchison, J.},
  year={2003},
  publisher={New Jersey: Reprinted by The Blackburn Press}
}

@article{stewart2011,
  title={{Managing the Essential Zeros in Quantitative Fatty Acid Signature Analysis}},
  author={Stewart, C. and Field, C.},
  journal={Journal of Agricultural, Biological, and Environmental Statistics},
  volume={16},
  number={1},
  pages={45--69},
  year={2011},
  publisher={Springer}
}

@article{stewart2017,
  title={An approach to measure distance between compositional diet estimates containing essential zeros},
  author={Stewart, C.},
  journal={Journal of Applied Statistics},
  volume={44},
  number={7},
  pages={1137--1152},
  year={2017},
  publisher={Taylor & Francis}
}

@article{scealy2011,
  title={Regression for compositional data by using distributions defined on the hypersphere},
  author={Scealy, J.L. and Welsh, A.H.},
  journal={Journal of the Royal Statistical Society. Series B},
  volume={73},
  number={3},
  pages={351--375},
  year={2011},
  publisher={Wiley Online Library}
}

@article{scealy2014,
  title={{Colours and cocktails: Compositional data analysis 2013 Lancaster lecture}},
  author={Scealy, JL and Welsh, AH},
  journal={Australian \& New Zealand Journal of Statistics},
  volume={56},
  number={2},
  pages={145--169},
  year={2014},
  publisher={Wiley Online Library}
}

@article{ait1989,
  title={Measures of location of compositional data sets},
  author={Aitchison, John},
  journal={Mathematical Geology},
  volume={21},
  number={7},
  pages={787--790},
  year={1989},
  publisher={Springer}
}

@inproceedings{tsagris2011,
  title={A data-based power transformation for compositional data},
  author={Tsagris, M.T. and Preston, S. and Wood, A.T.A.},
  booktitle={Proceedings of the 4rth Compositional Data Analysis Workshop, Girona, Spain},
  year={2011},
}

@article{tsagris2014,
  title = {{The $k-NN$ algorithm for compositional data: a revised approach with and without zero values present}},
  author = {Tsagris, Michail},
  journal = {Journal of Data Science},
  volume = {12},
  number = {3},
  pages={519--534},
  year={2014}
}

@inproceedings{tsagris2015a,
  title={A novel, divergence based, regression for compositional data},
  author={Tsagris, M.},
  booktitle={Proceedings of the 28th Panhellenic Statistics Conference, April, Athens, Greece},
  year={2015},
}

@article{tsagris2018b,
  title={{A Dirichlet regression model for compositional data with zeros}},
  author={Tsagris, Michail and Stewart, Connie},
  journal={Lobachevskii Journal of Mathematics},
  volume={39},
  number={3},
  pages={398--412},
  year={2018},
  publisher={Springer}
}

@article{martin2003b,
  title={Dealing with zeros and missing values in compositional data sets using nonparametric imputation},
  author={Mart{\'\i}n-Fern{\'a}ndez, Josep A and Barcel{\'o}-Vidal, Carles and Pawlowsky-Glahn, Vera},
  journal={Mathematical Geology},
  volume={35},
  number={3},
  pages={253--278},
  year={2003},
  publisher={Springer}
}

@article{otero2005,
  title={{Relative vs. absolute statistical analysis of compositions: a comparative study of surface waters of a Mediterranean river}},
  author={Otero, N and Tolosana-Delgado, R and Soler, A and Pawlowsky-Glahn, Vera and Canals, A},
  journal={Water Research},
  volume={39},
  number={7},
  pages={1404--1414},
  year={2005},
  publisher={Elsevier}
}

@article{pantazis2019,
  title={Gaussian asymptotic limits for the $\alpha$--transformation in the analysis of compositional data},
  author={Pantazis, Yannis and Tsagris, Michail and Wood, Andrew T.A.},
  journal={Sankhya A},
  volume={81},
  number={1},
  pages={63-82},
  year={2019},
  publisher={Springer}
}

@article{stewart2014,
  title={Testing for a change in diet using fatty acid signatures},
  author={Stewart, Connie and Iverson, Sara and Field, Christopher},
  journal={Environmental and Ecological Statistics},
  volume={21},
  number={4},
  pages={775--792},
  year={2014},
  publisher={Springer}
}

@article{tsagris2020,
  title={A folded model for compositional data analysis},
  author={Tsagris, Michail and Stewart, Connie},
  journal={Australian \& New Zealand Journal of Statistics},
  volume={62},
  number={2},
  pages={249--277},
  year={2020}
}

@article{hron2010,
  title={Imputation of missing values for compositional data using classical and robust methods},
  author={Hron, Karel and Templ, Matthias and Filzmoser, Peter},
  journal={Computational Statistics \& Data Analysis},
  volume={54},
  number={12},
  pages={3095--3107},
  year={2010},
  publisher={Elsevier}
}

@article{kendall2011,
  title={{Limit theorems for empirical Fr{\'e}chet means of independent and non-identically distributed manifold-valued random variables}},
  author={Kendall, Wilfrid S and Le, Huiling},
  journal={Brazilian Journal of Probability and Statistics},
  volume={25},
  number={3},
  pages={323--352},
  year={2011},
  publisher={Brazilian Statistical Association}
}

@article{palarea2008,
  title={{A modified EM alr-algorithm for replacing rounded zeros in compositional data sets}},
  author={Palarea-Albaladejo, Javier and Mart{\'\i}n-Fern{\'a}ndez, Josep Antoni},
  journal={Computers \& Geosciences},
  volume={34},
  number={8},
  pages={902--917},
  year={2008},
  publisher={Elsevier}
}

@article{endres2003,
  title={A new metric for probability distributions},
  author={Endres, Dominik Maria and Schindelin, Johannes E},
  journal={IEEE Transactions on Information Theory},
  volume={49},
  number={7},
  pages={1858--1860},
  year={2003},
  publisher={IEEE}
}

@article{osterreicher2003,
  title={A new class of metric divergences on probability spaces and its applicability in statistics},
  author={Osterreicher, Ferdinand and Vajda, Igor},
  journal={Annals of the Institute of Statistical Mathematics},
  volume={55},
  number={3},
  pages={639--653},
  year={2003},
  publisher={Kluwer Academic Publishers}
}

@article{ait1992,
  title={{On Criteria for Measure of Compositional Difference}},
  author={Aitchison, J.},
  journal={Mathematical Geology},
  volume={24},
  number={4},
  pages={365--379},
  year={1992},
  publisher={Springer}
}

@incollection{robcompositions2011,
    title = {{robCompositions: an R-package for robust statistical analysis of compositional data}},
    author = {Templ, Matthias and Hron, Karel and Filzmoser, Peter},
    publisher = {{John Wiley and Sons}},
    booktitle = {{Compositional Data Analysis: Theory and Applications}},
    editor={Vera Pawlowsky-Glahn and Antonella Buccianti},
    year = {2011},
    pages = {341--355},
}

@book{little2019,
  title={Statistical analysis with missing data},
  author={Little, Roderick JA and Rubin, Donald B},
  year={2019},
  publisher={John Wiley \& Sons}
}

@article{lin1991,
  title={Divergence measures based on the Shannon entropy},
  author={Lin, Jianhua},
  journal={IEEE Transactions on Information Theory},
  volume={37},
  number={1},
  pages={145--151},
  year={1991},
  publisher={IEEE}
}

@article{bear2016,
  title={A logistic normal mixture model for compositional data allowing essential zeros},
  author={Bear, John and Billheimer, Dean},
  journal={Austrian Journal of Statistics},
  volume={45},
  number={4},
  pages={3--23},
  year={2016}
}

@article{xavier2018,
  title={{Disaggregating statistical data at the field level: An entropy approach}},
  author={Xavier, Ant{\'o}nio and Freitas, Maria de Bel{\'e}m Costa and do Socorro Ros{\'a}rio, Maria and Fragoso, Rui},
  journal={Spatial Statistics},
  volume={23},
  pages={91--108},
  year={2018},
  publisher={Elsevier}
}

@article{lopez2020,
  title={{Reducing ignorance about who dies of what: research and innovation to strengthen CRVS systems}},
  author={Lopez, Alan D and McLaughlin, Deirdre and Richards, Nicola},
  journal={BMC Medicine},
  volume={18},
  number={58},
  year={2020},
  publisher={Springer}
}

@article{mattas2026,
  title={Using synthetic farm data to estimate individual nitrate leaching levels},
  author={Mattas, Konstantinos and Tsagris, Michail and Tzouvelekas, Vangelis},
  journal={American Journal of Agricultural Economics},
  year={2026},
  volume={108},
  number={1},
  pages={336--362},
  publisher={Wiley Online Library}
}

@book{molenberghs2007,
  title={{Missing Data in Clinical Studies}},
  author={Molenberghs, Geert and Kenward, Michael G.},
  year={2007},
  publisher={Wiley},
  address={Chichester}
}

@article{doove2014,
  author = {Doove, L. L. and Van Buuren, S. and Dusseldorp, E.},
  title = {Recursive partitioning for missing data imputation in the presence of interaction effects},
  journal = {Computational Statistics \& Data Analysis},
  volume = {72},
  pages = {92--104},
  year = {2014}
}

@article{shah2014,
  author = {Shah, A. D. and Bartlett, J. W. and Carpenter, J. and Nicholas, O. and Hemingway, H.},
  title = {{Comparison of random forest and parametric imputation models for imputing missing data using {MICE}: a {CALIBER} study}},
  journal = {American Journal of Epidemiology},
  volume = {179},
  number = {6},
  pages = {764--774},
  year = {2014}
}

@article{vanBuuren2006,
  author = {van Buuren, S. and Brand, J. P. L. and Groothuis-Oudshoorn, C. G. M. and Rubin, D. B.},
  title = {Fully conditional specification in multivariate imputation},
  journal = {Journal of Statistical Computation and Simulation},
  volume = {76},
  number = {12},
  pages = {1049--1064},
  year = {2006},
  doi = {10.1080/10629360600810434}
}

@article{vanBuuren2011,
  author = {van Buuren, S. and Groothuis-Oudshoorn, K.},
  title = {{mice: Multivariate imputation by chained equations in R}},
  journal = {Journal of Statistical Software},
  volume = {45},
  number = {3},
  pages = {1--67},
  year = {2011}
}

@article{xiaoqin2017,
  title={Imputation of missing values for compositional data based on random forest},
  author={XiaoQin, ZHANG and Yuying, CHENG},
  journal={Chinese Journal of Applied Probability and Statistics},
  volume={33},
  number={1},
  pages={102--110},
  year={2017}
}

@manual{tsagris2026,
 title = {{CompositionalNAimp: Missing Value Imputation with Compositional Data
}},
 author = {Michail Tsagris},
 year = {2026},
 note = {R package version 1.0},
 url = {https://CRAN.R-project.org/package=CompositionalNAimp}
}

\end{document}